\documentstyle[eqsecnum,aps,psfig]{revtex}

\begin{document}
\draft

\title{
	The $^{144}$Sm--$\alpha$ optical potential at astrophysically
	relevant energies derived from
	$^{144}$Sm($\alpha$,$\alpha$)$^{144}$Sm elastic scattering
}

\author{
	P.~Mohr, T.~Rauscher\footnote{APART--fellow, 
	present adress: Institut f\"ur theoretische Physik,
	Universit\"at Basel, Klingelbergstr.~82, CH--4056 Basel, Switzerland},
	and H.~Oberhummer
}

\address{
	Institut f\"ur Kernphysik, Technische Universit\"at Wien,
	Wiedner Hauptstra{\ss}e 8--10, A--1040 Wien, Austria
}

\author{
	Z.~M\'at\'e, Zs.~F\"ul\"op, and E.~Somorjai
}

\address{
	Institute of Nuclear Research of the Hungarian Academy of Sciences,
	PO Box 51, H-4001 Debrecen, Hungary
}

\author{
	M.~Jaeger\footnote{present address: Institut f\"ur Strahlenphysik,
	Universit\"at Stuttgart, Allmandring 3, D-70569 Stutt\-gart, Germany} 
	and G.~Staudt
}

\address{
	Physikalisches Institut, Universit\"at T\"ubingen,
	D-72076 T\"ubingen, Germany
}

\date{\today}

\maketitle

\begin{abstract}
For the determination of the $^{144}$Sm--$\alpha$ optical potential 
we measured the angular distribution of
$^{144}$Sm($\alpha$,$\alpha$)$^{144}$Sm scattering
at the energy $E_{\rm{lab}} = 20~{\rm{MeV}}$
with high accuracy. Using the known systematics of
$\alpha$--nucleus optical potentials we are able to derive
the $^{144}$Sm--$\alpha$ optical potential at the astrophysically
relevant energy $E_{\rm{c.m.}} = 9.5~{\rm{MeV}}$ 
with very limited uncertainties.
\end{abstract}




\section{Introduction}
\label{sec:intro}
In a detailed study of nucleosynthesis in Type II supernovae
Woosley and Howard proposed the so--called
$\gamma$--process, which is important for the production
of the samarium isotopes $^{144}$Sm and $^{146}$Sm \cite{woosley78}.
The existence of the $\gamma$--process was confirmed 
later \cite{rayet90}. Because of the $\alpha$ decay of
$^{146}$Sm ($T_{1/2} = 1.03 \times 10^8~{\rm{y}}$)
today one can find correlations between the
$^{142}$Nd/$^{144}$Nd ratio and the Sm/Nd ratio
in some meteorites \cite{prinz89} as a consequence
of the $\gamma$--process.
The nuclear reaction rates used in the network
calculations of the $\gamma$--process are relatively uncertain.
Especially, the reaction rate for the production of $^{144}$Sm
by the photodisintegration
reaction $^{148}$Gd($\gamma$,$\alpha$)$^{144}$Sm is uncertain by
a factor of 10 \cite{woosley90}.

The reaction rate for the reaction $^{148}$Gd($\gamma$,$\alpha$)$^{144}$Sm 
was derived from the $^{144}$Sm($\alpha$,$\gamma$)$^{148}$Gd reaction
cross section using a detailed--balance calculation
(see e.g.~\cite{woosley90,rauscher95}). Usually, reaction rates are given
at the relatively high temperatures of $2.5 \le T_9 \le 3.0$, or the
capture cross section at $E_{\rm{c.m.}} = 9.5~{\rm{MeV}}$ 
(corresponding to $T_9 = 2.8$)
is derived \cite{woosley78,woosley90,rauscher95,mann78,mohr94}.

Two ingredients enter into the Hauser--Feshbach calculation of the
$^{144}$Sm($\alpha$,$\gamma$)$^{148}$Gd capture cross section: transition
probabilities and the nuclear level density. The transition
probabilities were calculated using optical wave functions in an
equivalent square well (ESW) potential \cite{woosley78,woosley90},
a Woods--Saxon (WS) potential \cite{mann78,mohr94}, and
a folding potential \cite{rauscher95,mohr94}.

Calculations with ESW potentials are very sensitive
on a proper choice of the radius parameter $R$; two
calculations using $R = 8.01~{\rm{fm}}$ 
(Ref.~\cite{woosley90}, based on the ESW radius from Ref.~\cite{michaud70}) 
and $R = 8.75~{\rm{fm}}$ 
(Ref.~\cite{woosley78}, based on Ref.~\cite{truran72})
differ by a factor of 10 for the $^{144}$Sm($\alpha$,$\gamma$)$^{148}$Gd
cross section.
The WS potential and the folding potential 
in Refs.~\cite{rauscher95,mann78,mohr94} were taken
from global parametrizations of $\alpha$ optical potentials;
these results for the $^{144}$Sm($\alpha$,$\gamma$)$^{148}$Gd cross section
lie between the different ESW results. First experimental
results on the $^{144}$Sm($\alpha$,$\gamma$)$^{148}$Gd capture cross section
at somewhat higher energies lie at the lower end of the different 
calculations \cite{somorjai96}.

The aim of this work is to determine the optical potential
at the relevant energy $E_{\rm{c.m.}} = 9.5~{\rm{MeV}}$. In general, optical
potentials can be derived from elastic scattering angular
distributions. 
Recently, in a systematic study the energy and mass dependence
of $\alpha$--nucleus potentials was determined \cite{atzrott96}.
In that work $\alpha$ scattering on $^{144}$Sm was analyzed at higher
energies. An extrapolation to astrophysically relevant energies
is possible only with very limited accuracy.
At the energy $E_{\rm{c.m.}} = 9.5~{\rm{MeV}}$ 
the $^{144}$Sm($\alpha$,$\alpha$)$^{144}$Sm 
scattering cross section is given by almost pure
Rutherford scattering because of the height of the
Coulomb barrier of about 20 MeV. 
For a reliable determination of the optical potential
one has to increase the energy in the scattering experiment.
However, because of the energy
dependence of the optical potential, the energy should be
as close as possible to the astrophysically relevant energy 
$E_{\rm{c.m.}} = 9.5~{\rm{MeV}}$.
As a compromise we measured the $^{144}$Sm($\alpha$,$\alpha$)$^{144}$Sm 
angular distribution at $E_{\rm{lab}} = 20~{\rm{MeV}}$. 
At this energy the influence
of the nuclear potential on the angular distribution is measurable,
even though it is small.
Especially the determination of the shape of the optical potential
remains difficult even at this energy. For the real part this
problem vanishes because its shape is given by a folding procedure,
but the shape of the imaginary part has to be adjusted to the
experimental angular distribution.
For that reason the angular distribution
has to be determined with very high accuracy in the full
angular range.

The increase of the real part of the optical potential at energies close
to the Coulomb barrier is well-known \cite{satchler91}. 
Close to the Coulomb barrier the number
of open reaction channels changes strongly; this leads to a strong variation
of the imaginary potential, and the strength of the real potential is coupled
to the imaginary part by a dispersion relation.
A further influence on the potential strength coming from
antisymmetrization effects is indicated by microscopic calculations
which were performed for light nuclei.
This ``threshold anomaly'' (mainly so-called in heavy-ion scattering and 
fusion) was seen also in $\alpha$ scattering on many 
nuclei \cite{atzrott96,satchler91,mahaux86_1,mahaux86_2,michel95,abele93}.

\section{Experimental Setup and Data Analysis}
\label{sec:exp}
The experiment was performed at the cyclotron laboratory at ATOMKI,
Debrecen. We used the $78.8~{\rm{cm}}$ diameter scattering chamber 
which is described in detail in Ref.~\cite{mate89}. 
Here we will discuss only
those properties which are important for our experiment.

\subsection{Targets and Scattering Chamber}
\label{subsec:targets}
The samarium targets were produced by the reductive evaporation 
method \cite{westgaard66} at the target
laboratory at ATOMKI directly before the beamtime to avoid the
oxidation of the metallic samarium. A thin carbon foil (thickness
$d \approx 20~{\rm{\mu g/cm^2}}$) was used as backing.

During the experiment we used one enriched $^{144}$Sm and one natural Sm
target, and one carbon backing without samarium layer. The enrichment
in $^{144}$Sm was $(96.52 \pm 0.03)\%$. The thickness of the samarium
targets was determined by the energy loss
of the $\alpha$ particles
in the samarium layer. We compared the energy of $\alpha$ particles
elastically scattered on $^{12}$C in the carbon backing at
$\vartheta_{\rm{lab}} = 162^o$ using the pure carbon target as reference
and both samarium targets. The stopping powers $dE/dx$ 
of the $\alpha$ particles in samarium
at the relevant energies E = 20 MeV (incident $\alpha$) and E = 5.2 MeV
(backward scattered $\alpha$) were taken from Ref.~\cite{ziegler}.
The resulting thicknesses are
$d = 142~{\rm{\mu g/cm^2}}$ ($^{144}$Sm) and $d = 218~{\rm{\mu g/cm^2}}$
(natural Sm) with uncertainties of about 10\%. The pure carbon target
was used also for the angular calibration 
(see Sect.~\ref{subsec:angle}).

Additionally, two apertures were mounted on the target holder to check
the beam position and the size of the beam spot directly at the position of 
the target. 
The smaller aperture
had a width and height of 2 mm and 6 mm, respectively. 
This aperture was placed at the target position instead of the Sm target
before and after each variation of the beam current.
Because practically no
current could be measured on this aperture the width of the beam spot
was definitely smaller than 2 mm during the whole experiment, 
which is very important for the precise
determination of the scattering angle. In contrast, the relatively
poor determination of the height of the beam spot does not disturb
the claimed precision of the scattering angle (see Sect.~\ref{subsec:angle}).
Furthermore, the position of the beam on the target was continuously
controlled by two monitor detectors. No evidence was found for a change
of the position by determining the ratio of the count rates in both 
detectors (see Sect.~\ref{subsec:detectors}).

\subsection{Detectors}
\label{subsec:detectors}
For the measurement of the angular distribution we used 4 silicon
surface-barrier detectors with an active area $A = 50~{\rm{mm^2}}$
and thicknesses between $d = 300~{\rm{\mu m}}$ and $d = 1500~{\rm{\mu m}}$.
The detectors were mounted on an upper and a lower turntable, which can
be moved independently. On each turntable two detectors were
mounted at an angular distance of $10^o$. Directly in front of the
detectors apertures were placed with the dimensions 1.25 mm x 5.0 mm
(lower detectors) and 1.0 mm x 6.0 mm (upper detectors). Together with
the distance from the center of the scattering chamber $d = 195.6~{\rm{mm}}$
(lower detectors) and $d = 196.7~{\rm{mm}}$ (upper detectors)
this results in solid angles
from $\Delta \Omega = 1.63 \times 10^{-4}$ to 
$\Delta \Omega = 1.55 \times 10^{-4}$. The ratios of the solid angles of
the different detectors were determined by overlap measurements
with an accuracy much better than 1\%.

Additionally, two detectors were mounted at the wall of the
scattering chamber at a fixed angle of $\vartheta = 15^o$ (left and right side
relative to the beam direction). The solid angles of these detectors are
$\Delta \Omega = 8.10 \times 10^{-6}$. These detectors were used
as monitor detectors during the whole experiment.

The signals from all detectors were processed using charge-sensitive
preamplifiers (PA), which were mounted directly at the scattering chamber.
The output signal was further amplified by a main amplifier (MA). The bipolar
output of the MA was used by a Timing Single Channel Analyzer (TSCA)
to select signals with amplitudes between $\approx$ 1 V and 10 V, and
the unipolar output of the MA was gated with the TSCA signal using a
Linear Gate Stretcher (LGS). The LGS output was fed into an CAMAC ADC, and the
ADC data were stored in a corresponding CAMAC Histogramming Memory module.
The data acquisition was controlled by a standard PC with a CAMAC
interface using the program MCMAIN \cite{MCMAIN}. The dead time of the system
was determined by test pulses, which were fed into the test input
of each PA. It turned out that the deadtime
was negligible ($< 0.2\%$) except the runs at very forward angles.

The achieved energy resolution was better than 0.5\% corresponding
$\Delta E \le 100~{\rm{keV}}$ at $E_{\alpha} \approx 20~{\rm{MeV}}$.

\subsection{Angular Calibration}
\label{subsec:angle}
Because of the strong angular dependence of the scattering cross section
especially at forward angles the angular calibration was done very
carefully by two kinematic methods. The carbon backing contained some
hydrogen contamination. Therefore, we used the steep kinematics
of $^{1}$H($\alpha$,$\alpha$)$^{1}$H scattering at forward angles 
($10^o < \vartheta_{\rm{lab}} < 15^o$). We measured the energies of the
$\alpha$ particles scattered from $^{1}$H 
(note: two peaks with different $\alpha$ energies corresponding to
two center-of-mass scattering angles can be
found at one laboratory scattering angle, labelled $\pm$ in Eq.~\ref{eq:qpm})
and from $^{12}$C (ground state and $2^+$ state
at 4.44 MeV), and we determined the ratio
\begin{equation}
q^{\pm}(\vartheta) =
	\frac{ 
		E_{\alpha}(^{12}{\rm C}_{\rm{g.s.}}) 
		- E^{\pm}_{\alpha}(^{1}{\rm H}) 
	}
	{
		E_{\alpha}(^{12}{\rm C}_{\rm{g.s.}}) 
		- E_{\alpha}(^{12}{\rm C}_{2^+})
	}
\label{eq:qpm}
\end{equation}
from the experimentally measured energies and from a calculation
of the reaction kinematics. 
The angular offset is given by the mean value of the
differences in $\vartheta$ derived from all determined
values of $q^{\pm}_{\rm{exp}}$ and $q^{\pm}_{\rm{calc}}$.
The following results can be obtained from this procedure:
$\Delta \vartheta_{\rm{offset}} = -0.38^o \pm 0.02^o$ (lower detectors) and
$\Delta \vartheta_{\rm{offset}} = +0.32^o \pm 0.02^o$ (upper detectors).
(Note that these offsets cannot be a consequence of a beam spot which
is not exactly centered on the target; for the measured offset of
about $0.35^o$ the beam spot would have been outside the aperture
with a width of 2 mm, which can be placed at the position of the target,
see Sect.~\ref{subsec:targets}.)
The uncertainty of the adjustment of the angle is given by the
standard deviation of each single measurement:
$\Delta \vartheta_{\rm{adjust}} < \pm 0.1^o$ (all detectors).

\begin{figure}
\psfig{file=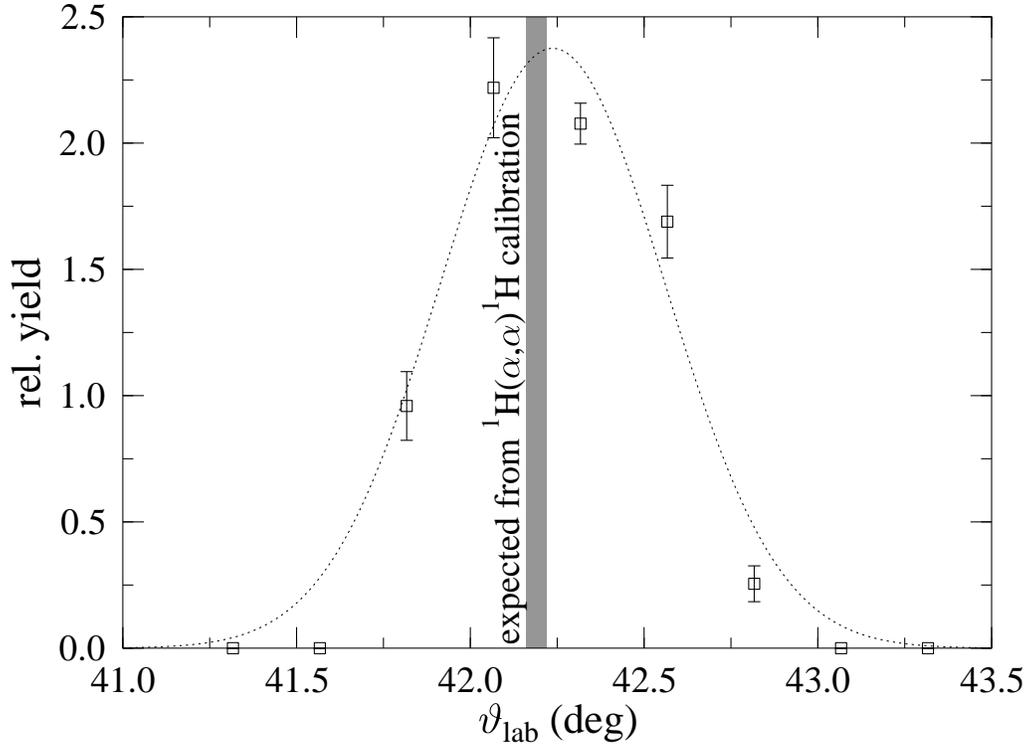,bbllx=50pt,bblly=50pt,bburx=550pt,bbury=400pt,clip=}
\caption{
Relative yield of $^{12}$C recoil nuclei in coincidence with elastically
scattered $\alpha$ particles. The shaded area shows the angle and the
uncertainty which is expected from the calibration using the steep
kinematics of $^{1}$H($\alpha$,$\alpha$)$^{1}$H. The dotted line is a
Gaussian fit to the experimental data points to guide the eye.
}
\label{fig:coin}
\end{figure}

In a second step we measured a kinematic coincidence between 
elastically scattered $\alpha$ particles and the corresponding $^{12}$C
recoil nuclei. One detector was placed at $\vartheta_{\rm{lab},\alpha} = 77^o$
(lower detector, left side relative to the beam axis), and the signals
from elastically scattered $\alpha$ particles on $^{12}$C were selected by an
additional TSCA. This TSCA output was used as gate for the signals from
another detector which was moved around the corresponding 
$^{12}$C recoil angle $\vartheta_{\rm{lab},recoil} = 42^o$ 
(upper detector, right side).
The maximum recoil count rate was found almost exactly at the expected
angle (see  Fig.~\ref{fig:coin}).

\subsection{Experimental Procedure, Uncertainties}
\label{subsec:exp}
\begin{figure}
\psfig{file=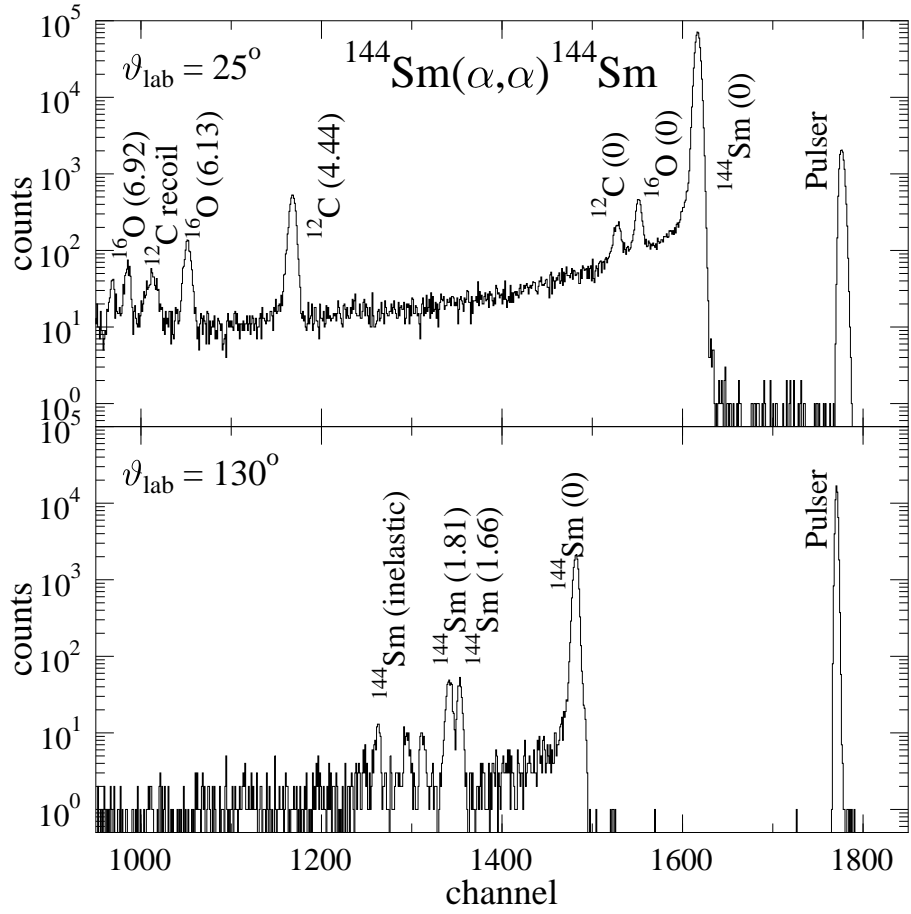,bbllx=50pt,bblly=50pt,bburx=550pt,bbury=600pt,clip=}
\caption{
Typical spectra of $^{144}$Sm($\alpha$,$\alpha$)$^{144}$Sm
at $\vartheta_{\rm{lab}} = 25^o$ (upper diagram) and
at $\vartheta_{\rm{lab}} = 130^o$ (lower diagram).
}
\label{fig:spec}
\end{figure}

With this setup we measured spectra from the $^{144}$Sm and the natural
Sm target at angles from $15^o$ to $172^o$ in steps of $1^o$
($\vartheta < 140^o$) and $2^o$ ($\vartheta > 140^o$). 
Two typical spectra measured at forward ($\vartheta = 25^o$) and
at backward ($\vartheta = 130^o$) angles are shown in Fig.~\ref{fig:spec}.
The measuring times were between some seconds (forward angles) and
several hours (backward angles),
and the corresponding beam currents
were between 30 nA and 600 nA $^4$He$^{2+}$ ions. The beam was stopped
in a Faraday cup roughly 2 m behind the scattering chamber, and the
current was measured by a current integrator.

For each run the scattering cross section was determined
relative to the monitor count rate:
\begin{equation}
\Bigl(\frac{d\sigma}{d\Omega}\Bigr)(\vartheta) =
	\Bigl(\frac{d\sigma}{d\Omega}\Bigr)_{\rm{Mon}}(\vartheta=15^o) \cdot
	\frac{N(\vartheta)}{N_{\rm{Mon}}(\vartheta=15^o)} \cdot
	\frac{\Delta \Omega_{\rm{Mon}}}{\Delta \Omega}
\end{equation}
and by assuming that the cross section at $\vartheta_{\rm{lab}} = 15^o$ 
is given
by pure Rutherford scattering. An absolute determination of the scattering
cross section from the accumulated charge and the target thickness
agrees within the quoted uncertainties with the relative determination
using the monitor detectors.

\begin{figure}
\psfig{file=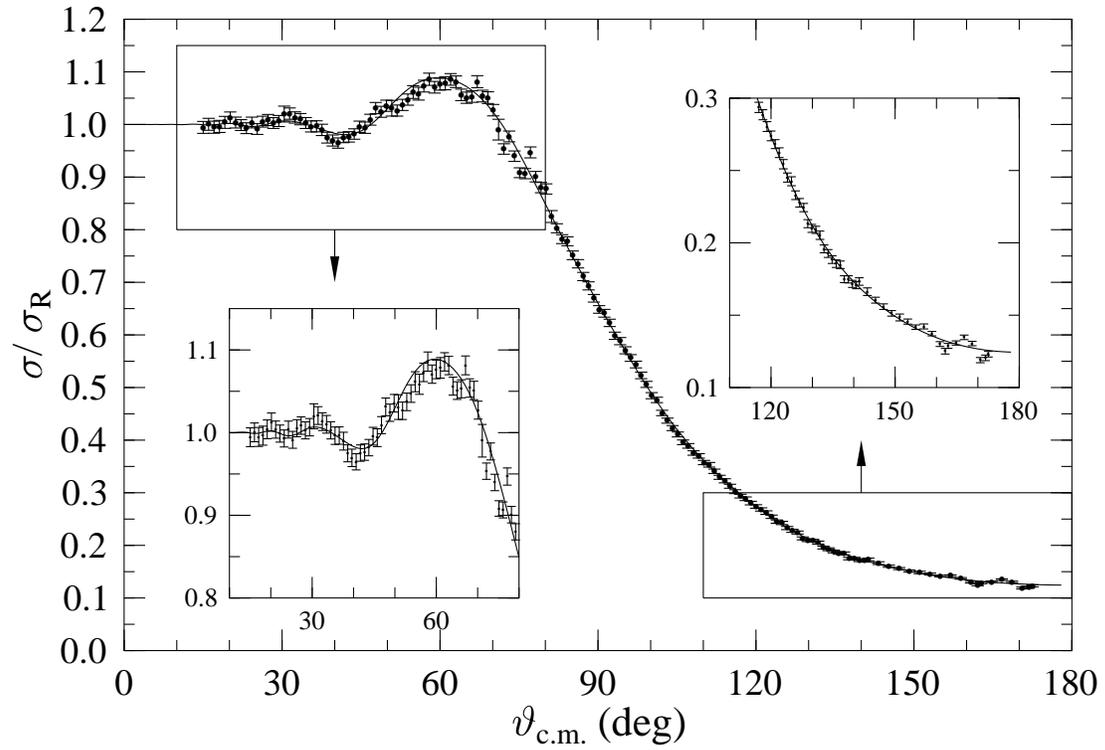,bbllx=50pt,bblly=50pt,bburx=550pt,bbury=600pt,clip=}
\caption{
Elastic scattering cross section of $^{144}$Sm($\alpha$,$\alpha$)$^{144}$Sm
normalized to the Rutherford cross section. The line is 
the result of an OM calculation using folding potentials and corresponds
to fit 1 of Tabs.~1 and 2
(see Sect.~III). 
The inserts show magnifications of the forward and backward angular range.
}
\label{fig:wq1}
\end{figure}

In the measured angular range the cross section 
covers more than 4 orders of magnitude.
The result of the experiment, normalized to the Rutherford cross section,
is shown in Fig.~\ref{fig:wq1}. The error bars
in Fig.~\ref{fig:wq1} contain statistical uncertainties ($< 1\%$ for
almost any angle) and systematic uncertainties coming from the
accuracy of the angular adjustment of the detectors and from contributions
of other samarium isotopes in the $^{144}$Sm target (chemical impurities of the
targets are negligible).

At forward angles the accuracy of the angular adjustment ($< \pm 0.1^o$)
leads to an uncertainty of about 1\% in the determination of the cross
section, at backward angles this uncertainty 
in the cross section practically disappears.
In contrast, at forward angles the elastic scattering cross section of
all samarium isotopes is very close to the Rutherford cross section, however
at backward angles the cross section measured with the natural samarium target
is about 20\% to 30\% smaller than the $^{144}$Sm scattering cross section.
Because of the high enrichment (96.52\%) of the $^{144}$Sm target the resulting
uncertainty remains smaller than 1\% even at backward angles. Therefore,
we renounced of a correction of the $^{144}$Sm scattering cross section.

The resulting high accuracy of this experiment is better than 2\%
(including both statistical and systematic uncertainties)
even for the very backward angles where the cross section is more than
4 orders of magnitude smaller than at the forward angles measured in this
experiment.

\section{Optical Model Analysis}
\label{sec:OM}
The theoretical analysis of the scattering data was performed in the
framework of the optical model (OM). 
The complex optical potential is given by
\begin{equation}
U(r) = V_{\rm C}(r) + V(r) + iW(r)
\end{equation}
where $V_{\rm C}(r)$ is the Coulomb potential, and $V(r)$ resp.~$W(r)$ are the
real and the imaginary part of the nuclear potential, respectively.

The real part of the optical potential was calculated by a
double--folding procedure:
\begin{equation}
V_{\rm f}(r) = 
  \int \int \rho_{\rm P}(r_{\rm P}) \, \rho_{\rm T}(r_{\rm T}) \,
  v_{\rm eff}(E,\rho = \rho_{\rm P} + \rho_{\rm T}, 
	s = |\vec{r}+\vec{r_{\rm P}}-\vec{r_{\rm T}}|) \,
  d^3r_{\rm P} \, d^3r_{\rm T}
\label{eq:fold}
\end{equation}
where $\rho_{\rm P}$, $\rho_{\rm T}$ are the densities of projectile and target,
respectively,
and $v_{\rm eff}$ is the effective nucleon-nucleon interaction taken in the
well-established DDM3Y parametrization \cite{satchler79,kobos84}.
Details about the folding procedure can be found in 
Refs.~\cite{abele93,atzrott96}, the folding integral 
in Eq.~\ref{eq:fold}
was calculated using the code DFOLD \cite{DFOLD}.
The strength of the folding potential is adjusted by the usual
strength parameter $\lambda$ with $\lambda \approx 1.2 - 1.3$.

The densities of the $\alpha$ particle and the $^{144}$Sm nucleus
were derived from the experimentally known charge
density distributions \cite{devries87}, assuming identical
proton and neutron distributions. For $N \approx Z$ nuclei
up to $^{90}$Zr ($Z = 40$, $N = 50$) this assumption works well,
however, in the case of $^{208}$Pb ($Z = 82$, $N = 126$) a theoretically
derived neutron distribution and the experimental proton
distribution had to be used to obtain a good description
of the elastic scattering angular distribution
\cite{atzrott96}. 
To take 
the possibility into account that the proton and neutron distributions are
not identical in the nucleus $^{144}$Sm ($Z = 62$, $N = 82$)
a scaling parameter $w$ for the width of the potential
was introduced, which is very close to unity. 
The resulting real part of the optical potential
is given by:
\begin{equation}
V(r) = \lambda \cdot V_{\rm f}(r/w)
\end{equation}

For a comparison of different potentials we use the
integral parameters volume integral per interacting
nucleon pair $J_R$ and the root-mean-square (rms) radius
$r_{\rm{rms,R}}$, which are given by:
\begin{eqnarray}
J_R & = &
	\frac{1}{A_{\rm P} A_{\rm T}} \, \int V(r) \, d^3r \\
r_{\rm{rms,R}} & = &
	\left[ \frac{\int V(r) \, r^2 \, d^3r}
			{\int V(r) \, d^3r} \right] ^{1/2}
\end{eqnarray}
for the real part of the potential $V(r)$ and corresponding equations
hold for $W(r)$.
The values for the folding potential $V_{\rm f}$ 
at $E_{\rm{lab}} = 20~{\rm{MeV}}$ (with $\lambda = w = 1$)
are $J_R = 260.41~{\rm{MeV fm^3}}$ and $r_{\rm{rms,R}} = 5.573~{\rm{fm}}$.

The Coulomb potential is taken in the usual form of a 
homogeneously charged sphere where the Coulomb radius $R_{\rm C}$
was chosen identically with the rms radius of the
folding potential $V_{\rm f}$: $R_{\rm C} = r_{\rm{rms,R}} = 5.573~{\rm{fm}}$.

For the imaginary part of the potential different parametrizations
were chosen: the usual Woods--Saxon (WS) potential
\begin{equation}
W_{\rm{WS}}(r) = W_0 \cdot ( 1 + \exp{(r-R)/a} )^{-p}
\end{equation}
where $R$ is usually given by $R = R_0 \cdot A_{\rm T}^{1/3}$,
and a series of Fourier-Bessel (FB) functions
\begin{equation}
W_{\rm{FB}}(r) = \sum_{k=1}^{n} 
	a_k \sin{(k \pi r/R_{\rm{FB}})} / (k \pi r/R_{\rm{FB}})
\end{equation}
with the cutoff radius $R_{\rm{FB}}$.

A fitting procedure was used to minimize the deviation $\chi^2$ between
the experimental and the calculated cross section:
\begin{equation}
\chi^2 = \sum_{i=1}^N 
	\Bigl( \frac{\sigma_{\rm{exp},i}(\vartheta) 
		- \sigma_{\rm{calc},i}(\vartheta)}
	{\Delta \sigma_{\rm{exp},i}(\vartheta)} \Bigr)^2
\end{equation}
The calculations were performed using the code A0 \cite{A0}.

The parameters of the imaginary part, the potential strength parameter
$\lambda$ and the width parameter $w$ of the real part, and the absolute value
of the angular distribution were adjusted to the experimental data
by the fitting procedure. The ratio $r$ of the calculated to the experimental
absolute value of the angular distribution is very close to 1 in all fits:
$r = 1.006 \pm 0.002$. Therefore, we renormalized the measured
angular distribution by this ratio $r$. This renormalization of 0.6\% lies well
within our experimental uncertainties.

\begin{figure}
\psfig{file=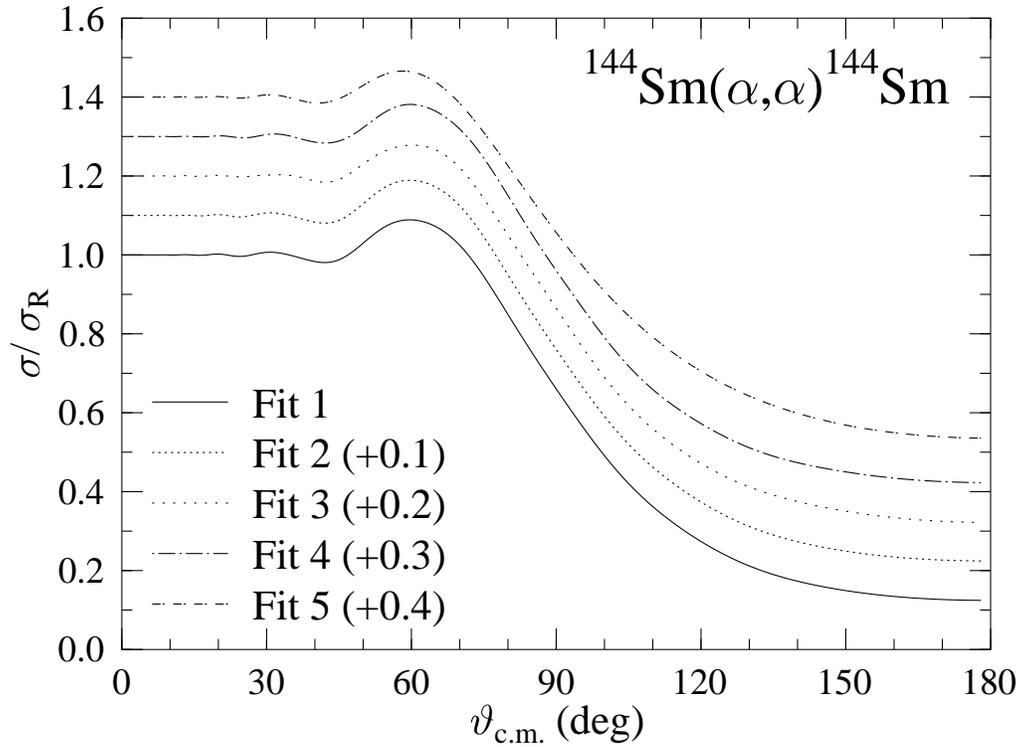,bbllx=50pt,bblly=50pt,bburx=550pt,bbury=600pt,clip=}
\caption{
Calculated elastic scattering cross section 
of $^{144}$Sm($\alpha$,$\alpha$)$^{144}$Sm
normalized to the Rutherford cross section. The lines are 
the result of different OM calculations using folding potentials
(fits 1--4, which have almost the same $\chi^2/F$) 
and using a standard WS potential (fit 5) from
Ref.~\protect\cite{mann78}. The parameters are shown in
Tabs.~1 and 2.
}
\label{fig:wq2}
\end{figure}

Four parametrizations of the imaginary part of the potential were used
to determine the influence of the type of imaginary parametrization 
on the resulting volume integrals.
They are labelled by 1, 2, 3 and 4 in Tabs.~\ref{tab:tab1} and \ref{tab:tab2}.
Labels 1 and 2 correspond to WS potentials with p=1 and p=2, respectively. 
For the parametrizations 3 and 4 FB-functions were used with 5 and 8
FB coefficients, respectively.
The results are listed in Tabs.~\ref{tab:tab1} and \ref{tab:tab2},
the calculated angular distributions are shown in Figs.~\ref{fig:wq1}
and \ref{fig:wq2}. In addition, 
also the angular distribution derived from a
``standard'' WS potential is shown,
which was used in previous calculations of the
$^{144}$Sm($\alpha$,$\gamma$)$^{148}$Gd capture cross section 
\cite{mann78,mohr94}.
The fits 1--4 look very similar; the $\chi^2/F$ varies from 1.74 to 1.82.
However, the fits 3 and 4 using FB functions in the imaginary part
show a slightly oscillating imaginary potential; fit 4 even shows a small
unphysical region where the imaginary part becomes positive.

Of course, the $\chi^2/F$ of fit 5 is inferior compared to fits 1--4
because the WS parameters
taken from the study of Ref.~\cite{mann78} were not adjusted to the
experimental angular distribution.

\section{Discussion}
\label{sec:disc}

\subsection{Ambiguities of the optical potential}
\label{subsec:ambi}
We extracted a definite optical potential from the experimental
elastic scattering data on $^{144}$Sm($\alpha$,$\alpha$)$^{144}$Sm
at $E_{\rm{lab}} = 20~{\rm{MeV}}$. First of all, very accurately
measured scattering data are necessary for this determination,
and furthermore a definite solution has to be selected from
several potentials which describe the data almost identically.
These problems result from discrete ambiguities (the so--called
``family problem'') and from continuous ambiguities.

The problem of continuous ambiguities is reduced to a great
extent by the use of folding potentials because the shape
of the folding potential is better fixed compared to standard
potentials of WS type. The width parameter of the
folding potential, which was introduced in Sect.~\ref{sec:OM}
should remain very close to unity; otherwise the parameters extracted
from this calculation are not very reliable.

\begin{figure}
\psfig{file=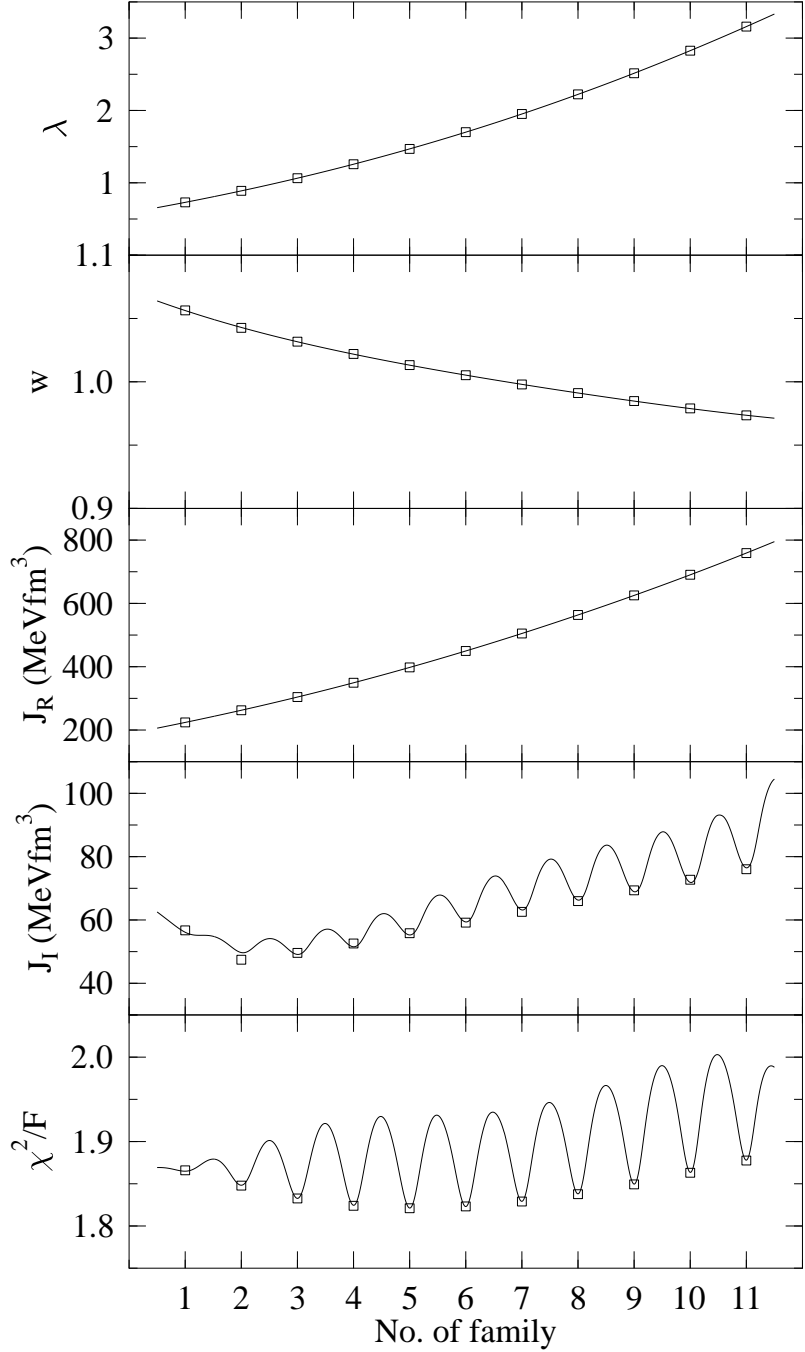,height=550pt}
\caption{
Parameters of different families (numbered 1--11)
of optical potentials derived from the
$^{144}$Sm($\alpha$,$\alpha$)$^{144}$Sm scattering data:
strength parameter $\lambda$, width parameter $w$, 
volume integrals of the real and imaginary part $J_R$ resp.~$J_I$,
and the deviation per degree of freedom $\chi^2/F$
(from top to bottom). The data points are results of fits
of real and imaginary part using different starting values,
the lines are the result of an
interpolation in $\lambda$ and $w$ to adjust the real part
of the potential to $V(r = 10.61~{\rm{fm}}) = -0.54~{\rm{MeV}}$,
and the imaginary part of the potential was again adjusted to the
experimental data. The deepest minima in $\chi^2/F$ are obtained
for the families 4, 5, and 6.
}
\label{fig:family}
\end{figure}

The ``family problem'' is illustrated in Fig.~\ref{fig:family}.
Because of the reasons mentioned above we used the simplest parametrization
of the imaginary potential, the standard WS type with $p = 1$
as employed in fit 1.
If one now varies continuously the depth of the real part of the optical
potential (i.~e.~the strength parameter $\lambda$) and adjusts
as well the width parameter $w$ and the parameters of the imaginary part
to the experimental data, then one obtains a continuous variation
of $w$ and $J_R$, but oscillations in $\chi^2/F$ correlated with
oscillations in $J_I$. Each (local) minimum 
in $\chi^2/F$ -- shown as data points
in Fig.~\ref{fig:family} -- corresponds to one family of the optical
potential.
Obviously, the deepest minima in $\chi^2/F$ can be
obtained for the families 4, 5, and 6 (see Fig.~\ref{fig:family}).
This restriction on the number of the family is confirmed by
the behaviour of the width parameter $w$, which should be
very close to 1.
Fit.~1 in Sec.~\ref{sec:OM} (see Figs.~\ref{fig:wq1}, \ref{fig:wq2})
can be found here as family 4. 

\begin{figure}
\psfig{file=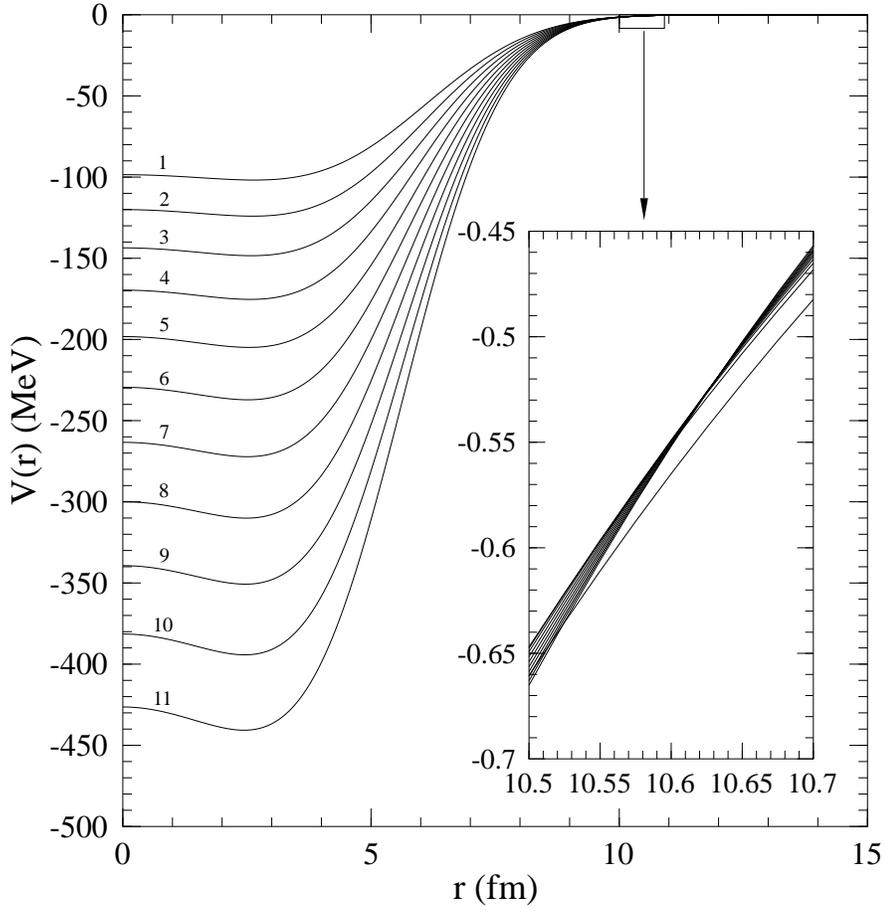,bbllx=50pt,bblly=50pt,bburx=550pt,bbury=600pt,clip=}
\caption{
Real part of different families of optical potentials (labelled 1--11)
derived from the
$^{144}$Sm($\alpha$,$\alpha$)$^{144}$Sm scattering data:
in the insert one can see that all potentials have the
same depth at the radius $r = 10.61~{\rm{fm}}$:
$V(r = 10.61~{\rm{fm}}) = -0.54~{\rm{MeV}}$. For family 1 (dashed line)
one finds an exception but for the region of
this family $\chi^2/F$ does not show a certain minimum
(see Fig.~\protect\ref{fig:family}).}
\label{fig:pot}
\end{figure}

A more detailed analysis of the resulting real potentials shows
that all potentials 
have the same depth at the radius $r = 10.61~{\rm{fm}}$:
$V(r = 10.61~{\rm{fm}}) = -0.54~{\rm{MeV}}$.
This result is shown in Fig.~\ref{fig:pot}. 
(An exception was found for family 1, but for the region of
this family $\chi^2/F$ does not show a certain minimum,
see Fig.~\ref{fig:family}.)
A similar result was already found by
Badawy {\it et al.} \cite{badawy78} by 
analyzing excitation functions of $\alpha$ scattering measured 
close to $\vartheta = 180^o$ at energies 
$10~{\rm{MeV}} \le E_{\rm{lab}} \le 20~{\rm{MeV}}$:
These authors stated that
``at energies near the Coulomb barrier the $\alpha$--particle
scattering data are fitted by any Woods--Saxon potential
whose depth at $r = R_0$ is 0.2 MeV''.
For $^{144}$Sm they derived $R_0 = 11.04 \pm 0.02~{\rm{fm}}$.
However, in contradiction to that statement we point out
that one has to choose one of the discrete values of the real
volume integrals $J_R$ determined by the minima in $\chi^2/F$
(families 1--11).
Choosing e.g.\ a strength parameter
$\lambda = 1.35$ (which is exactly between families 4 and 5)
and adjusting $w$ so that $V(r = 10.61~{\rm{fm}}) = -0.54~{\rm{MeV}}$
($w = 1.018$) results in a significantly worse description
of the experimental data ($\chi^2/F = 1.92$ compared to
$\chi^2/F = 1.82$ for families 4 and 5). Of course, this 
discrimination is only possible if very accurately measured
scattering data are available!

\begin{figure}
\psfig{file=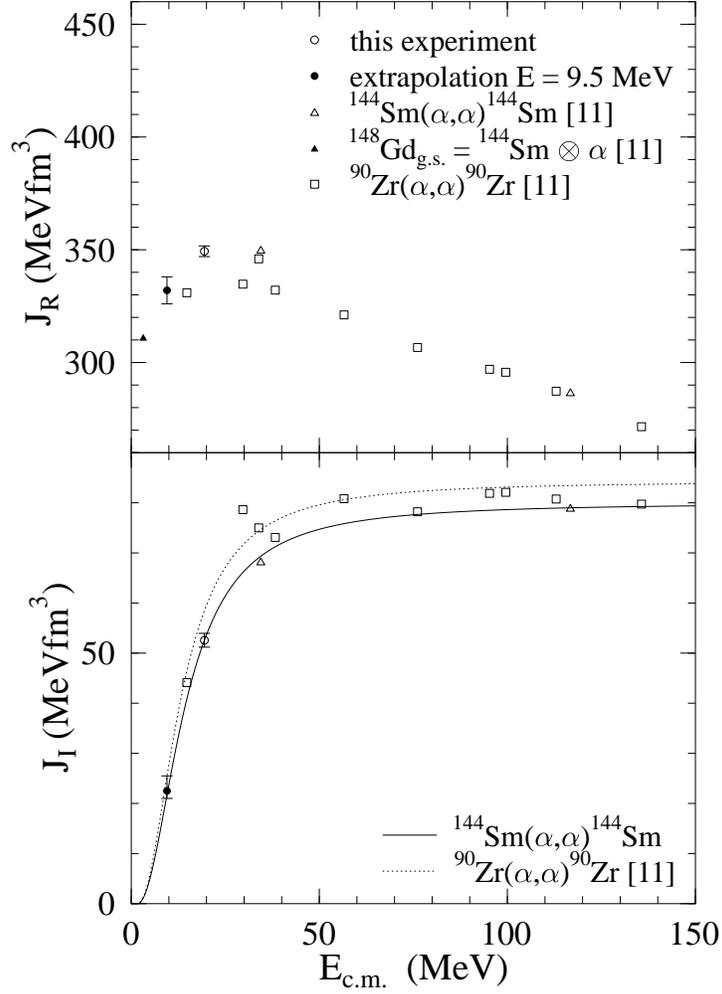,bbllx=50pt,bblly=50pt,bburx=550pt,bbury=600pt,clip=}
\caption{
Volume integrals of the real (upper) and imaginary part 
(lower diagram) of the optical potential
derived from $^{144}$Sm($\alpha$,$\alpha$)$^{144}$Sm scattering.
For comparison the volume integrals
derived from $^{90}$Zr($\alpha$,$\alpha$)$^{90}$Zr scattering 
\protect\cite{atzrott96} were added.
The lines in the lower diagram show the results of BR parametrizations
of the imaginary part (see text).}
\label{fig:vol}
\end{figure}

The ``family problem'' usually can be solved at higher energies.
The $^{144}$Sm($\alpha$,$\alpha$)$^{144}$Sm scattering data
at $E_{\rm{lab}} = 120~{\rm{MeV}}$ \cite{ichi87} can be described
well by a calculation using a folding potential with
$J_R = 286.8~{\rm{MeVfm^3}}$ \cite{atzrott96},
and a similar volume integral was found in
Ref.~\cite{ichi87} using WS potentials.
Together with the systematic behaviour
of folding potentials for nuclei with $A \ge 90$ \cite{atzrott96} 
which can be described by the interplay of the energy dependence of
the NN interaction $v_{\rm eff}$ and the effect of the so--called
``threshold anomaly'' \cite{satchler91,abele93}
one expects volume integrals of about $J_R \approx 330 - 340~{\rm{MeVfm^3}}$
at the low energies analyzed in this work
(see Fig.~\ref{fig:vol}, upper part and Sect.~\ref{subsec:extra}).
>From this point of view we can decide that family 4
corresponds to the volume integral
$J_R (E_{\rm{lab}} = 120~{\rm{MeV}}) = 286.8~{\rm{MeVfm^3}}$. 

There is a further confirmation for family 4.
The ground state wave function of $^{148}$Gd$_{\rm{g.s.}}$ = $^{144}$Sm
$\otimes$ $\alpha$ can be calculated \cite{atzrott96}.
The number of nodes $N$ and 
the angular momentum $L$ of the $\alpha$ particle outside
the $^{144}$Sm core are related to the oscillator quantum number $Q$
by 
\begin{equation}
Q = 2N + L = \sum_{i=1}^4 (2n_i + l_i) = \sum_{i=1}^4 q_i .
\end{equation}
Using $Q = 18$ (corresponding to 
$q_i = 5$ oscillator quanta for each neutron in the $2f_{7/2}$ shell and 
$q_i = 4$ oscillator quanta for each proton in the $2d_{5/2}$ shell,
see e.g.~\cite{mk})
one has to adjust the folding potential
strength to reproduce the binding energy
of the $^{148}$Gd ground state. One obtains
$\lambda = 1.159$ and $J_R = 311.2~{\rm{MeVfm^3}}$
at E = +3.2 MeV (the nucleus $^{148}$Gd decays by $\alpha$ emission)
which fits into the known systematic behaviour of $\alpha$-nucleus
volume integrals \cite{atzrott96}.

For all these reasons we chose the calculations
corresponding to family 4 with
$J_R \approx 349~{\rm{MeVfm^3}}$ and $J_I \approx 52.5~{\rm{MeVfm^3}}$
(see Sect.~\ref{sec:OM}) obtained with the standard WS parametrization
of the imaginary part (fit 1 in Tabs.~\ref{tab:tab1}
and \ref{tab:tab2}).

\subsection{Extrapolation to $E_{\rm{c.m.}} = 9.5~{\rm{MeV}}$}
\label{subsec:extra}
For the calculation of the 
$^{144}$Sm($\alpha$,$\gamma$)$^{148}$Gd 
reaction cross section at the astrophysically relevant
energy $E_{\rm{c.m.}} = 9.5~{\rm{MeV}}$ one has to determine the optical
potential at that energy. The following methods
were applied to extract the real and imaginary
part of the potential.

In a first step the folding potential in the real part
was calculated at the energy $E_{\rm{c.m.}} = 9.5~{\rm{MeV}}$ (this is
necessary because of the energy dependence of the
interaction $v_{\rm eff}$, see Eq.~\ref{eq:fold}).
Second, the width parameter $w$ was taken from fit 1: $w = 1.022$.
Third, because of the rise of
the volume integrals $J_R$ at low energies
which is about 
$\Delta J_R / \Delta E \approx 1 - 2~{\rm{MeVfm^3/MeV}}$
\cite{atzrott96}
we adjusted the parameter $\lambda$ to obtain a
volume integral of 
$J_R(E_{\rm{c.m.}} = 9.5~{\rm{MeV}}) = 334 \pm 6~{\rm{MeVfm^3}}$:
$\lambda = 1.1965 \pm 0.0216$. The uncertainties of
$J_R$ and $\lambda$ were estimated from the
uncertainties of $\Delta J_R / \Delta E$ and
$J_R(E_{\rm{lab}} = 20~{\rm{MeV}})$.

The volume integral of the imaginary part
can be parametrized according to Brown and Rho (BR)
\cite{brown81}:
\begin{equation}
J_I(E_{\rm{c.m.}}) = \left\{ \begin{array}{rll}
	& \multicolumn{1}{c}{0} 
		& {\mbox{for~}} E_{\rm{c.m.}} \le E_0 \\
	& J_0 \cdot \frac{(E_{\rm{c.m.}} - E_0)^2}
		{(E_{\rm{c.m.}} - E_0)^2 + \Delta^2}
		& {\mbox{for~}} E_{\rm{c.m.}} > E_0 \\
	\end{array} \right.
\label{eq:br}
\end{equation}
with the excitation energy $E_0$ of the first excited
state, and the saturation parameter $J_0$ and the
rise parameter $\Delta$, which are adjusted to the
experimentally derived values.

>From the $^{144}$Sm scattering data at $E_{\rm{lab}} = 20~{\rm{MeV}}$ and
$E_{\rm{lab}} = 120~{\rm{MeV}}$ \cite{ichi87} one obtains
$J_0 = 79.98~{\rm{MeVfm^3}}$ and $\Delta = 12.84~{\rm{MeV}}$.
The excitation energy of the first excited
$2^+$ state in $^{144}$Sm is $E_0 = 1.660~{\rm{MeV}}$.
This leads to a volume integral at $E_{\rm{c.m.}} = 9.5~{\rm{MeV}}$ of
$J_I(E_{\rm{c.m.}} = 9.5~{\rm{MeV}}) = 21.7~{\rm{MeVfm^3}}$.
Because of the weak mass dependence of the imaginary
part for heavy nuclei with a magic neutron or proton
number (see Fig.~\ref{fig:vol}, lower part)
and because the BR parametrization is not very well
defined by 2 data points (and 2 parameters to be adjusted!)
we also used the well--defined BR parametrization for
$^{90}$Zr taken from Ref.~\cite{atzrott96}:
$J_0 = 84.3~{\rm{MeVfm^3}}$ and $\Delta = 11.8~{\rm{MeV}}$.
This leads to 
$J_I(E_{\rm{c.m.}} = 9.5~{\rm{MeV}}) = 25.5 \pm 0.3~{\rm{MeVfm^3}}$
using either $E_0 = 1.66~{\rm{MeV}}$ adjusted to $^{144}$Sm
or $E_0 = 1.78~{\rm{MeV}}$ adjusted to $^{90}$Zr.
Combining the values derived from the BR parametrizations
of $^{144}$Sm and $^{90}$Zr, we adopt a volume integral of 
$J_I(E_{\rm{c.m.}} = 9.5~{\rm{MeV}}) = 22.5^{+3.0}_{-1.5}~{\rm{MeVfm^3}}$.

The shape of the imaginary potential is not as
certain as its volume integral.
Several parametrizations lead to almost
identical fits at $E_{\rm{lab}} = 20~{\rm{MeV}}$. For our extrapolation
we used the geometry of the potential derived in fit 1
(see Tabs.~\ref{tab:tab1} and \ref{tab:tab2}), 
because the shape of the imaginary potential should not
change dramatically from $E_{\rm{lab}} = 20~{\rm{MeV}}$ to 
$E_{\rm{c.m.}} = 9.5~{\rm{MeV}}$,
and we adjusted the depth $W_0 = 4.55^{+0.61}_{-0.30}~{\rm{MeV}}$ 
to obtain a potential with the correct imaginary 
volume integral.

\subsection{$^{144}$Sm($\alpha$,$\gamma$)$^{148}$Gd and the
$^{146}$Sm/$^{144}$Sm production ratio}
\label{subsec:alphagamma}
Reaction rates for the reaction $^{144}$Sm($\alpha$,$\gamma$)$^{148}$Gd
at $T_9 = 2.5$ and $T_9 = 3.0$ are listed in 
Tab.~\ref{tab:tab3}. 
To determine the influence of the optical potential on the
$^{144}$Sm($\alpha$,$\gamma$)$^{148}$Gd cross section
the calculations of Refs.~\cite{rauscher95,mohr94} were repeated
only changing the optical potentials.
Compared to a previous calculation
using a folding potential with parameters derived from a
global systematics \cite{rauscher95,mohr94}
the reaction rate is reduced by a factor of about 1.5.
The optical potential is well--defined by the scattering data.
With the optical potential determined in this work the calculated and the
preliminary experimental $^{144}$Sm($\alpha$,$\gamma$)$^{148}$Gd capture
cross section agree reasonably well.

The reaction rate does not depend strongly on the chosen family
because the scattering data are reproduced quite well from calculations
using potentials of families 3, 4, and 5.
The capture cross section decreases (increases) by about 11\% (10\%)
using family 3 (5) instead of family 4. A similar increase of about 8\%
compared to family 4 is obtained when one uses a potential between families
4 and 5 as mentioned in Sect.~\ref{subsec:ambi} because of the somewhat larger
imaginary part of the potential.

The influence of the reaction rate on the $^{146}$Sm/$^{144}$Sm production
ratio was analyzed in Ref.~\cite{woosley90}. Our reduced reaction rate
lies between cases C and D of Tab.~I in Ref.~\cite{woosley90}
corresponding to a production ratio of about 0.3 which is well between
the experimentally derived limits of 0.1 to 0.7 \cite{prinz89}.

\section{Conclusion}
\label{sec:conclusion}
The elastic scattering cross section 
$^{144}$Sm($\alpha$,$\alpha$)$^{144}$Sm 
was measured at $E_{\rm{lab}} = 20~{\rm{MeV}}$ with very high accuracy.
A definite optical potential could be derived
from the experimental data and by taking into
account the systematics of $\alpha$--nucleus optical
potentials of Ref.~\cite{atzrott96}. Again using this
systematics, an extrapolation of the optical potential
from $E_{\rm{lab}} = 20~{\rm{MeV}}$ to the astrophysically relevant energy
$E_{\rm{c.m.}} = 9.5~{\rm{MeV}}$ was possible with very limited uncertainties.
The $^{144}$Sm($\alpha$,$\gamma$)$^{148}$Gd cross section is reduced
by a factor of about 1.5 compared to a previous folding potential
calculation, it still lies between the two
ESW calculations differing by a factor of 10. The uncertainty of the
$^{144}$Sm($\alpha$,$\gamma$)$^{148}$Gd cross section coming from 
continuous and discrete ambiguities of the optical
potential is reduced by a great amount. The use
of systematic folding potentials is highly recommended
for the analysis of low--energy $\alpha$ scattering
and $\alpha$ capture reactions
because of the reduced number of free parameters
compared to previous calculations using WS potentials,
resp.~because of the reduced uncertainties in
the radius parameter compared to ESW calculations.

\acknowledgments
We would like to thank the cyclotron team of ATOMKI for the
excellent beam during the experiment. Two of us (P.~M., M.~J.)
gratefully acknowledge the kind hospitality at ATOMKI.
This work was supported by the 
Austrian--Hungarian Exchange (projects B43, OTKA T016638),
Fonds zur F\"orderung der wissenschaftlichen Forschung 
(FWF project S7307--AST), and
Deutsche Forschungsgemeinschaft (DFG project Mo739).

\newpage

\begin{table}
\caption{
	Potential parameters of the imaginary part of
	the optical potential
	derived from the angular distribution of
	$^{144}$Sm($\alpha$,$\alpha$)$^{144}$Sm
	at $E_{\rm{lab}} = 20~{\rm{MeV}}$.
}
{
\begin{tabular}{cccccccccc}
fit     & $W_0~({\rm{MeV}})$    
	& $R_0~({\rm{fm}})$     
	& $a~({\rm{fm}})$       
	& $p$   
	& & & & & \\
\hline
%
1       & 10.64 & 1.6758        & 0.1680        & 1     & & & & & \\
2       & 10.70 & 1.7132        & 0.2265        & 2     & & & & & \\
\hline
fit     & $R_{\rm{FB}}~({\rm{fm}})$
	& $a_1$ & $a_2$ & $a_3$ & $a_4$
	& $a_5$ & $a_6$ & $a_7$ & $a_8$ \\
\hline
%
3       & 15.0
	& -11.65        & -16.66        & 6.09          & 21.31
	& 10.79         & -             & -             & - \\
4       & 15.0
	& -9.50         & -6.50         & 19.42         & 14.84
	& -23.51        & -35.83        & -14.88        & -1.06 \\
\hline
5 \tablenotemark[1]     &
\multicolumn{9}{c}{$V_0 = 185~{\rm{MeV}}$, $W_0 = 25~{\rm{MeV}}$,
	$R_{0,R} = R_{0,I} = 1.4~{\rm{fm}}$, $a_R = a_I = 0.52~{\rm{fm}}$}
\end{tabular}
\tablenotetext[1]{Ref.~\protect\cite{mann78}}
}
\label{tab:tab1}
\end{table}

\begin{table}
\caption{
	Integral potential parameters $J$ and $r_{\rm{rms}}$
	of the real and imaginary part of the optical potential
	derived from the angular distribution of
	$^{144}$Sm($\alpha$,$\alpha$)$^{144}$Sm
	at $E_{\rm{lab}} = 20~{\rm{MeV}}$.
}
{
\begin{tabular}{cccccccc}
fit     & $\lambda$     & $w$   
	& $J_R$ & $r_{\rm{rms,R}}$
	& $J_I$ & $r_{\rm{rms,I}}$
	& $\chi^2/F$ \\
No.     &               &       
	& $({\rm{MeV fm^3}})$   & $({\rm{fm}})$
	& $({\rm{MeV fm^3}})$   & $({\rm{fm}})$
	& \\
\hline
%
1       & 1.2568                & 1.0220
	& 349.34                & 5.6961        & 52.57 & 6.8311        & 1.823 \\
2       & 1.2580                & 1.0216
	& 349.29                & 5.6940        & 52.41 & 6.8142        & 1.823 \\
3       & 1.3425                & 0.9976
	& 347.04                & 5.5600        & 57.49 & 6.0756        & 1.807 \\
4       & 1.2771                & 1.0067
	& 339.31                & 5.6110        & 51.30 & 6.8868        & 1.738 \\
%
\hline
5 \tablenotemark[1]     &
\multicolumn{2}{c}{Woods--Saxon}
	& 557.59                & 6.0026        & 75.35 & 6.0026        & 41.0
\end{tabular}
}
\tablenotetext[1]{Ref.~\protect\cite{mann78}}
\label{tab:tab2}
\end{table}

\begin{table}
\caption{
	Reaction rates of $^{144}$Sm($\alpha$,$\gamma$)$^{148}$Gd
	at temperatures of $T_9 = 2.5$ and $T_9 = 3.0$.
}
{
\begin{tabular}{ccccc}
potential
	& potential     & ($\alpha$,$\gamma$)
	& \multicolumn{2}{c}
	{$N_A \cdot < \sigma v >~({\rm{cm^3 \, s^{-1} \, mole^{-1}}})$} \\
	& from Ref.     & from Ref.
	& $T_9 = 2.5$
	& $T_9 = 3.0$ \\
\hline
ESW, R = 8.01 fm
	& \protect\cite{michaud70}
	& \protect\cite{woosley90}
	& 3.72 $\times$ 10$^{-16}$      & 2.58 $\times$ 10$^{-13}$ \\
ESW, R = 8.75 fm 
	& \protect\cite{truran72}
	& \protect\cite{woosley78}
	& 3.75 $\times$ 10$^{-15}$      & 2.35 $\times$ 10$^{-12}$ \\
WS 
	& \protect\cite{mann78}
	& \protect\cite{mann78,mohr94}
	& 1.95 $\times$ 10$^{-15}$      & 1.22 $\times$ 10$^{-12}$ \\
Folding, $\lambda = 1.159$ 
	& \protect\cite{rauscher95,mohr94}
	& \protect\cite{rauscher95,mohr94}
	& 1.27 $\times$ 10$^{-15}$      & 7.56 $\times$ 10$^{-13}$ \\
Folding 
	& this work
	& this work
	& 7.91 $\times$ 10$^{-16}$      & 5.63 $\times$ 10$^{-13}$ \\
\end{tabular}
}
\label{tab:tab3}
\end{table}

\newpage

\clearpage

\clearpage

\clearpage

\clearpage

\clearpage

\clearpage

\end{document}